\documentclass{aa}
\usepackage{graphicx}
\usepackage{txfonts}
\usepackage{natbib}
\bibpunct{(}{)}{;}{a}{}{,}
\begin{document}
   \title{\textit{INTEGRAL} observations of AGN in the Galactic Plane}


   \author{S. Soldi
          \inst{1,}\inst{2}
          \and V. Beckmann
	  \inst{3,}\inst{4}
	  \and L. Bassani
	  \inst{5}
	  \and T. J.-L. Courvoisier
          \inst{1,}\inst{2}
	  \and R. Landi
	  \inst{5,}\inst{6}
	  \and A. Malizia
	  \inst{5}
	  \and A. J. Dean
	  \inst{7}
	  \and A. De Rosa
	  \inst{8}	  
	  \and A. C. Fabian
	  \inst{9}
	  \and R. Walter
	  \inst{1,}\inst{2}
	  }

   \offprints{Simona.Soldi@obs.unige.ch}

   \institute{\textit{INTEGRAL} Science Data Centre, 
              Chemin d'\'Ecogia 16, 1290 Versoix, Switzerland
         \and Observatoire de Gen\`eve, 51 chemin des 
	      Mailletes, 1290 Sauverny, Switzerland 
         \and NASA Goddard Space Flight Center, Exploration of the Universe Division, 
	      Greenbelt, MD 20771, USA
	 \and Joint Center for Astrophysics, Department of Physics, University of Maryland, 
	      Baltimore County, MD 21250, USA
	 \and IASF, CNR/INAF, via Gobetti 101, 40129 Bologna, Italy
	 \and Dipartimento di Fisica, Universita` di Bologna, 
	      Viale C. Berti Pichat 6/2, I-40127 Bologna, Italy
	 \and School of Physics and Astronomy, University of Southampton, Highfield, Southampton, 
	      SO17 1BJ, UK	      
	 \and IASF, CNR/INAF, Via Fosso del Cavaliere 100, I-00133 Rome, Italy  
	 \and Institute of Astronomy, Madingley Road, Cambridge CB3 0HA, UK	 	 
	      }

   \date{Received 21 July 2005 / Accepted 22 August 2005}

   \abstract{We present results on approximately one year of \textit{INTEGRAL} observations of six AGN detected during 
   the regular scans of the Galactic Plane. The sample is composed by five Seyfert 2 objects ({\mbox MCG --05--23--16}, 
   {\mbox NGC 4945}, the Circinus galaxy, {\mbox NGC 6300}, {\mbox ESO 103--G35}) and the radio galaxy Centaurus~A. 
   The continuum emission of each of these sources is well represented by a highly absorbed ($N_{\rm H}>10^{22} \, \rm cm^{-2}$) 
   power law, with average spectral index 
   $\Gamma = 1.9 \pm 0.3$. A high energy exponential cut-off at $E_c \sim 50 \rm \, keV$ is required to fit the spectrum of the 
   Circinus galaxy,
   whereas a lower limit of 130 keV has been found for {\mbox NGC 4945} and no cut-off has been detected for {\mbox NGC 6300}
   in the energy range covered by these \textit{INTEGRAL} data.
   The flux of Centaurus~A was found to vary by a factor of $\sim 2$ in 10 months, showing 
   a spectral change between the high and low state, which can be modelled equally well by a change in the absorption ($N_{\rm H}$ from 
   17 to $33 \times 10^{22} \rm \, cm^{-2}$) or by the presence of a cut-off at $\ga$ 120 keV in the low state spectrum.
   A comparison with recently reprocessed \textit{BeppoSAX}/PDS data shows a general agreement with \textit{INTEGRAL} results.
   The high energy cut-off in the hard X-ray spectra appears to be a common but not universal characteristic of Seyfert 2 
   and to span a wide range of energies.   
   \keywords{Galaxies: active -- Galaxies: Seyfert -- Gamma rays: observations -- X-rays: galaxies 
             }
   }   

   \maketitle
%

\section{Introduction}

   In radio quiet Active Galactic Nuclei (AGN), most of the power supplied by the central black hole
   is emitted in the form of X-rays and $\gamma$-rays. This emission
   comes from the regions close to the nucleus and its study enables us to 
   constrain the geometry and the state of the matter in the heart of an AGN. 
   The emission in the soft X-ray domain is rather well known 
   (up to $\la$ 10 keV). The spectrum is in most cases well reproduced by an absorbed 
   power law, a Compton reflection component and a Fe $\rm K_{\alpha}$ fluorescence line.
   Many questions are however still open for the hard X-rays/soft $\gamma$-ray 
   range. 
   For many objects, a high-energy cut-off is required to reproduce
   the data between 60 and 300 keV, but the shape and the energy of 
   this feature is not yet well known (\citealt{zdziarski95}, \citealt{risaliti02},
   \citealt{deluit03}).
   
   Our knowledge of the hard X-ray sky above 10 keV comes from observations
   carried out by many instruments, like \textit{Granat}/SIGMA \citep{paul91}, 
   \textit{CGRO}/BATSE \citep{fishman92}, \textit{CGRO}/OSSE \citep{johnson93},
   \textit{BeppoSAX}/PDS \citep{frontera97} and \textit{RXTE}/HEXTE \citep{rothschild98}. \\
   The fine spectroscopy with imaging and accurate positioning of
   the sources makes \textit{INTEGRAL} (INTErnational Gamma-Ray Astrophysics Laboratory,
   \citealt{winkler03a})
   a suitable instrument to study the hard X-ray emission of AGN 
   also in a dense sky region like the Galactic Plane.    
   In the first year of Core Program observations, \textit{INTEGRAL} has detected a number of AGN.
   Among those are objects already known in the 20--100 keV band, 
   while others are new hard X-ray discoveries ({\citealt{bassani04},
   \citealt{masetti04}). 
   
   In this paper, we present a preliminary study of the hard X-ray emission for 6 AGN 
   detected by \textit{INTEGRAL} in its first year of observations and 
   compare our results to previous ones based on \textit{BeppoSAX}/PDS observations. 
   In Section~\ref{sample} we discuss our sample selection and in Section~\ref{analysis}
   we detail the data reduction and analysis of \textit{INTEGRAL} and \textit{BeppoSAX} 
   observations. 
   A variability study of these objects and the spectral analysis of the \textit{INTEGRAL} and \textit{BeppoSAX}/PDS
   data are reported in Sections~\ref{var} and \ref{fit}, respectively. 
   The results are discussed in Section~\ref{discussion} and conclusions reported in Section~\ref{conclusions}.
%

\section{Sample Selection}\label{sample}
 
   The sources in this sample were selected from among objects detected during
   the Core Program (CP herein after; \citealt{winkler03b}) observations performed by \textit{INTEGRAL}. 
   From February to October 2003 \textit{INTEGRAL}
   CP observations led to the detection of hard X-ray emission from 10 previously known 
   AGN: {\mbox MCG --05--23--16}, {\mbox NGC 4945}, Centaurus~A, Circinus galaxy, 
   {\mbox NGC 6300}, {\mbox GRS 1734--292}, {\mbox PKS 1830--211}, {\mbox ESO 103--G35}, {\mbox NGC 6814}, {\mbox Cygnus A} \citep{bassani04}.
   Among these, some were poorly observed in the hard X-ray domain by previous missions ({\mbox GRS 1734--292}, {\mbox PKS 1830--211}). 
   
   \textit{INTEGRAL} results for {\mbox GRS 1734--292} and {\mbox PKS 1830--211} have been already presented
   by \citet{sazonov04} and \citet{derosa05}, respectively, while an analysis of {\mbox NGC 6814}
   is in preparation. Cygnus A has not been 
   included in our sample due to likely contamination from the cluster emission surrounding the AGN.

   We present in this work the remaining 6 AGN: {\mbox MCG --05--23--16}, {\mbox NGC 4945}, Centaurus~A, 
   Circinus galaxy, {\mbox NGC 6300}, {\mbox ESO 103--G35}, which were previously studied by different 
   missions, in particular by \textit{BeppoSAX}, allowing us to compare the \textit{INTEGRAL} results
   with previous ones. 
   The objects in this sample (Table~\ref{src}) are at small cosmological distances 
   (averaged redshift of $z =$ 0.005) and are all Seyfert 2 galaxies according to their NED classification, 
   although Centaurus~A (a Narrow Line Radio Galaxy) shows a complex spectral energy distribution which makes
   a unique classification difficult. 

\begin{table}
\hspace{-0.3cm}
\begin{minipage}[t]{\columnwidth}
\caption{The sample}
\label{src}
\renewcommand{\footnoterule}{} 
\centering 
\begin{tabular}{c c c c c c}
\hline\hline                
Name & R.A.\footnote{J2000 coordinates.} & Decl.$^a$ & z & Type & $N_{\rm H} \rm \, gal$ \\  
     & (deg) & (deg) &  &  & $(\rm 10^{20} \, cm^{-2})$ \\    
\hline                        
   MCG --05--23--16 & 146.9 & --30.9 & 0.0083 & Sey2 & 8.0  \\	 
   NGC 4945         & 196.4 & --49.5 & 0.0019 & Sey2 & 15.7 \\
   Centaurus A      & 201.4 & --43.0 & 0.0018 & RG   & 8.6   \\ 
   Circinus galaxy  & 213.3 & --65.3 & 0.0015 & Sey2 & 55.6  \\ 
   NGC 6300         & 259.2 & --62.8 & 0.0037 & Sey2 & 9.4   \\ 
   ESO 103--G35     & 279.6 & --65.4 & 0.0133 & Sey2 & 7.7   \\ 
   \hline                                 
\end{tabular}
\end{minipage}
\end{table}
%

\section{Observations and Data Analysis}\label{analysis}

\subsection{\textit{INTEGRAL}}

   The \textit{INTEGRAL} mission is dedicated to the spectroscopy
   and fine imaging of sources in the energy range 15 keV -- 10 MeV, with 
   its principal instruments, the imager IBIS \citep{ubertini03} and the spectrometer SPI 
   \citep{vedrenne03}.
   Simultaneous observations are performed by the X-ray monitor JEM-X (3--35 keV, 
   \citealt{lund03}) and 
   the optical monitor OMC (V band, \citealt{mas-hesse03}).
   The IBIS telescope is composed by two independent detectors
   optimised for low (ISGRI, 15--1000 keV, \citealt{lebrun03}) and high (PICsIT, 0.175--10 MeV,
   \citealt{dicocco03}) energies. 
     
   The large field of view of IBIS ($9^o \times 9^o$ fully coded), the high
   angular resolution (12') and the typical on-axis sensitivity of 1 mCrab at 100 keV
   (3$\sigma$, $10^6$s) are characteristics optimised for a regular scan of the
   Galactic Plane \citep{winkler03b} in order to monitor the known sources and to discover new
   hard X-ray sources. 
   This survey, with a deep exposure of the Galactic Centre and pointed observations,
   constitutes the Core Program of the guaranteed time which involved
   $\sim$ 9.3 Ms of the observing time available during the first 
   year of the mission. 
     
   Each source of this sample was first detected \citep{bassani04} during the \textit{INTEGRAL} CP observations
   performed between 2003 February 28 (revolution 46) and October 10 (revolution 120),     
   in the search for significant excesses in the mosaic sky image (a weigthed mean
   of all single pointing images) of the IBIS/ISGRI instrument.
   In order to perform a more detailed analysis, we used additional data 
   including those that were public at the time of this study, specifically all the data in revolutions
   1--136 (October 2002--November 2003) and 142--149 (December 2003--January 2004). 
   For objects like Centaurus~A and the Circinus galaxy this substantially improved the results, thanks to 
   data from pointed observations. 
   In Table 2 the log of the \textit{INTEGRAL} observations are reported togheter with the ISGRI significance and count rates 
   of the sample sources.
   
   Analysis of data collected by ISGRI, JEM-X and SPI instruments has been performed, 
   selecting the observations made respectively within $10^\circ$, $5^\circ$ and $10^\circ$ 
   radius of each source. The extraction radius for JEM-X is smaller because of the smaller fully coded 
   field of view of $\sim 5^\circ$ radius.
   Imaging, spectra and timing analysis were performed using the version 4.2 of the
   ISDC's Offline Science Analysis (OSA) package \citep{courvoisier03a}. 
         
   Because of the faint nature of the sources in the sample, we chose to extract the ISGRI spectra 
   for {\mbox NGC 4945}, the Circinus galaxy, {\mbox NGC 6300} and {\mbox ESO 103--G35} from the count rate and variance 
   mosaic images at the position of the source, which in all cases
   corresponds to the brightest pixel in the 30--50 keV band. This band 
   provides the best compromise between maximising the signal-to-noise ratio 
   and having a good removal of imaging artefacts \citep{bird04}. 
   We built the spectra in 10 energy bands from 20 to 600 keV and then rebinned them
   in order to have at least 30 counts in each bin
   to apply the $\chi^2$ minimization technique. 
   Only the flux of Centaurus~A was high enough to allow the spectral extraction
   with the OSA standard method and to use a finer 
   binning (2 keV width). These data were then rebinned with the same criterium as described above. 
   The low detection significance of {\mbox MCG --05--23--16}
   does not allow spectral extraction. 
   Based on Crab calibration studies, we added a systematic error of 3\% only to the ISGRI spectra of the three 
   brigthest sources ({\mbox NGC 4945}, Centaurus~A and the Circinus galaxy), as for the other objects statistical
   errors would be dominant anyway.
   
   The ISGRI light curves for the three brightest objects were created with the standard
   software with time bin of 3000 seconds in the 20--40 keV and 40--60 keV bands. 
   A finer binning of 100 s was used 
   in order to perform a deeper analysis of only some specific periods. 
     
   Because of the smaller field of view (FoV) of JEM-X, the exposure times 
   for the X-ray monitor are much lower than for the IBIS and SPI observations.
   {\mbox MCG --05--23--16} was never in the field of view of JEM-X and {\mbox NGC 4945}, {\mbox NGC 6300} and 
   {\mbox ESO 103--G35} were not detected.
   A low significance ($2.7 \, \sigma$ in the 3--10 keV band, count rate of 0.3 $\pm$ 0.1 $\rm s^{-1}$ in the 3-35 keV) 
   for Circinus has been obtained 
   with 17 ks of data, allowing only an estimate 
   of the flux level, while spectral extraction between 3--35 keV has been performed for Centaurus~A,
   thanks to a highly significant detection of $\sim 84 \, \sigma$  for an exposure of 113 ks (count rate of 
   3.24 $\pm$ 0.04 $\rm s^{-1}$ in the 3-35 keV band). 
   JEM-X spectra of Centaurus~A have been extracted with the standard software for each pointing 
   and then a weighted average has been constructed to obtain spectra along longer periods. 
   
   The SPI analysis was done using the specific analysis software \citep{diehl03} including 
   version 9.2 of the reconstruction software SPIROS \citep{skinner03} which is based on 
   the ``Iterative Removal of Sources'' technique \citep{hammersley92}. 
   {\mbox NGC 4945}, Centaurus~A and Circinus have been detected by SPI, with count rates of
   (5.1 $\pm$ 0.7) $\times 10^{-3} \rm \, s^{-1}$ (20-200 keV), (1.64 $\pm$ 0.07) $\times 10^{-2} \rm \, s^{-1}$ (20-200 keV)
   and (3.3 $\pm$ 0.4) $\times 10^{-3} \rm \, s^{-1}$ (20-100 keV), respectively and  
   SPI spectra have been extracted for these three sources.
   Only $3 \sigma$ upper limits of 1.5, 0.6 and 1.3 $\times 10^{-10} \rm \, erg \, cm^{-2} s^{-1}$ 
   have been obtained in the 20--40 keV band for {\mbox MCG --05--23--16}, {\mbox NGC 6300} and
   {\mbox ESO 103--G35}, respectively, which are consistent with the ISGRI fluxes in the same energy
   band (1.0, 0.3 and 0.6$\times 10^{-10} \rm \, erg \, cm^{-2} s^{-1}$).
    
   The spectral analysis was performed using the {\sc XSPEC 11.3.1}
   software package \citep{arnaud96}. In the following, all quoted errors correspond to 90\% 
   confidence interval for one interesting parameter ($\Delta\chi^{2}=2.71$).
\begin{table*}
\begin{minipage}[t]{\columnwidth}
\caption{\textit{INTEGRAL} and \textit{BeppoSAX} observations}
\label{obs}
\renewcommand{\footnoterule}{}  
\centering 
\begin{tabular}{c c c c c c c c c c}
\hline\hline   \noalign{\smallskip}
     &  \multicolumn{4}{c}{\textit{INTEGRAL}} & & \multicolumn{4}{c}{\textit{BeppoSAX}}  \\
\noalign{\smallskip}     
\cline{2-5} \cline{7-10}
\noalign{\smallskip}     
Name & Obs Date & Exp$^{\rm a}$ & $\sigma$$^{\rm a}$ & Count rate$^{\rm a}$ &
& Obs Date & Exp$^{\rm b}$ & $\sigma$$^{\rm b}$ & Count rate$^{\rm b}$  \\     
     &      & (ks) &  & $(\rm s^{-1})$ &  & & (ks)  & &   $(\rm s^{-1})$  \\   
\hline                        
\noalign{\smallskip}     
   MCG --05--23--16 & Jul 2003            & 2   & 2.4   & 1.8  $\pm$ 1.1  & & Apr 24, 1998 & 34  & 36  & 1.59 $\pm$ 0.04 \\   
   NGC 4945        & Jan 2003 -- Jan 2004 & 276 & 40.2  & 2.38 $\pm$ 0.07 & & Jul 1, 1999  & 44  & 81  & 2.59 $\pm$ 0.03 \\
   Centaurus A     & Mar 2003 -- Jan 2004 & 404 & 210.8 & 9.86 $\pm$ 0.05 & & 1997--2000   & 90  & 196 & 5.58 $\pm$ 0.03 \\ 
   Circinus galaxy & Jan 2003 -- Jan 2004 & 589 & 57.2  & 2.29 $\pm$ 0.05 & & 1998--2001   & 138 & 92  & 1.93 $\pm$ 0.02 \\ 
   NGC 6300        & Mar 2003 -- Oct 2003 & 173 & 7.0   & 0.7 $\pm$ 0.1   & & Aug 28, 1999 & 42  & 37  & 1.25 $\pm$ 0.04 \\ 
   ESO 103--G35    & Mar 2003 -- Oct 2003 & 36  & 4.2   & 2.5 $\pm$ 0.3   & & 1996--1997   & 30  & 15  & 0.81 $\pm$ 0.05 \\ 
\hline       \noalign{\smallskip}                            
\end{tabular}
\end{minipage}
\begin{list}{}{}
\footnotetext \item \small{$^{\rm a}$ Exp: ISGRI exposure time not corrected for the off-axis vignetting effects;  }\,\,
\footnotetext \item \small{$\sigma$: ISGRI detection significance in the 20--100 keV band;} \,\, 
\footnotetext \item \small{Count rate: ISGRI background substracted 
                           count rate in the 20--300 keV band.} \\
\footnotetext \item \small{$^{\rm b}$ Exp: PDS effective exposure time;  }\,\,
\footnotetext \item \small{$\sigma$: PDS detection significance in the 20--100 keV band;} \,\, 
\footnotetext \item \small{Count rate: PDS background substracted 
                           count rate in the 15--200 keV band.} \\
\end{list}
\end{table*}

\subsection{\textit{BeppoSAX}}

   All the PDS data, publicly available through the \textit{BeppoSAX} archive have been
   analysed for the 6 AGN of the sample (Table~\ref{obs}). 
   For three of these sources (MCG--05--23--16, {\mbox NGC 4945}, {\mbox NGC 6300}) only one observation has been performed. 
   For the other three objects (Centaurus~A, the Circinus galaxy, {\mbox ESO 103--G35}) the summed spectrum of 
   several observations has been used (5, 2 and 2 observations, respectively) because no significant variation 
   of the flux and of the spectral parameters has been detected during the \textit{BeppoSAX} observations.      
   Spectral analysis has been performed in the range 15--150 or 15--200 keV with 10 energy bins. 
   
   The PDS spectra were extracted using the \emph{XAS} v2.1
   package \citep{chiappetti97} which provides smaller error bars\footnote{On~behalf~of~the~PDS~group,~see
   ftp://ftp.tesre.bo.cnr.it in the directory
   /pub/sax/doc/software\_docs/xas\_vs\_saxdas.ps.} than 
   \emph{SAXDAS}, i.e. the standard package for the reduction and analysis of \textit{BeppoSAX}/PDS data.
   It is also important to emphasize that a significant  
   improvement in the signal-to-noise ratio and a more reliable check of the 
   background fields are obtained using \emph{XAS}.
   Source visibility windows were selected following the criteria
   of no Earth occultation and high voltage stability during the
   exposure. In addition, the observations closest to the South
   Atlantic Anomaly were excluded from the analysis.
   Since most of the sources in our sample are rather faint in the PDS band, 
   it was necessary to carefully check the background subtraction by taking advantage 
   of the rocking technique \citep{frontera97}.
   When two units of the collimator are pointing ON source, the other two units are
   pointing towards one of the two OFF positions. 
   The OFF spectra are then used as background in the
   computation of the spectrum of the target source.
  
   Comparing the two offset fields, we found the presence of two contaminating
   sources in the -OFF fields of the Circinus galaxy observation: a $4 \,\sigma$ excess 
   at $RA = 14^{\rm h} 42' 50.7"$, $DEC = -63^\circ 47' 27.5"$ (J2000.0) in the first measurement and 
   a $14 \,\sigma$ excess at $RA = 14^{\rm h} 11' 34.0"$, $DEC = -61^\circ 48' 09.0"$ in the second exposure.
   The error associated to the above positions is quite large (around $1^{\circ}$) thus 
   making the search for likely X-ray counterparts a difficult task. Nevertheless, 
   the first object  may  be associated 
   to RCW86 ({\mbox SNR 315.0--02.3}), a supernova remnant, while the second
   one is probably the HMXRB system {\mbox 4U 1416--62}. 
        
   Also in the observation of {\mbox ESO 103--G35} we found a contaminating source in the 
   +OFF background field: the $5 \,\sigma$ excess at $RA = 18^{\rm h} 47' 03"$, $DEC = -62^\circ 05' 40.0"$,
   also detected in the XTE Slew Survey is likely the Seyfert 1 galaxy {\mbox ESO 140--43}.
   In both cases, in order to extract the 
   uncontaminated source spectra of both the Circinus galaxy and {\mbox ESO 103--G35}, we excluded 
   the contaminated fields
   and considered only the uncontaminated measurements in the computation of the background
   for these two sources. 
 

\section{Source Variability}\label{var}

   For the three brightest sources in this sample, i.e. {\mbox NGC 4945}, Centaurus~A and Circinus, we extracted 
   the light curves from \textit{INTEGRAL}/ISGRI data to study the variability over a period of about one year. \\
   The pointed observations of the Circinus galaxy took place in the period 6--18 July 2003, while between January and June 2003
   only a few data were collected with about monthly recurrence. Data from 4 pointings are available in the period
   December 24, 2003 -- January 2, 2004.
   {\mbox NGC 4945} observations are concentrated in the two periods March 7--9, 2003 and January 2--4, 2004.
   A few pointings executed in January, June and July 2003 have also been analysed. 
     
   The apparent increase of the source count rates observed at the end of 
   some revolutions (rev. 90 for Circinus and 48 for {\mbox NGC 4945}) is due to the increased contribution of 
   the background which is generally higher in the \textit{INTEGRAL} instruments when the satellite approaches the perigee passage
   due to radiation belt effects.
   Some modulation on time scales of hours in the light curve of {\mbox NGC 4945} was revealed to be due to an anticorrelation between 
   the apparent flux and the off-axis position of the source. 
   For these reasons it is difficult to study the intrinsic variability of so weak sources on time scales of hours. 
   We therefore rebinned the light curves in order to have one flux value per revolution and we excluded those revolutions in which the source
   was observed for less than 20 ks. 
   
   The light curve of the Circinus galaxy is represented by three data points in   
   the period 6--18 July 2003. A significant variability of 10\% within 2 days has been 
   detected in the 20-40 keV range. When fitted with a constant, the light curve gives $\chi^2 = 10$ with 2 degrees of freedom, 
   resulting in a probability that random fluctuations will produce a poorer reduced chi-squared of $P_{\chi} = 0.007$. 
   No significant variations have been seen in the 40-60 keV band ($\chi^2 = 1.1$, $P_{\chi} = 0.6$). 
   
   Two flux points in March 7--9, 2003 and one in January 2--4, 2004 constitute the light curve of NGC 4945, which presents a 20\% 
   decrease in flux in the 20--40 keV range within two days in March 2003. The fit with a constant gives $\chi^2 = 18$ (2 d.o.f.)
   with a significant null hypothesis probability of $P_{\chi} = 10^{-4}$ and only a marginal variability has been detected in the 40--60 keV band 
   ($\chi^2 = 6$, $P_{\chi} = 0.05$).
   
   A larger variation has been observed in the flux of Centaurus~A.
   The ISGRI light curve shows a decrease by a factor of $\sim 2$ in the count rate level in both the 20--40  
   and 40--60 keV ranges between the beginning (March 7--9, 2003) 
   and the end (January 2--4, 2004) of the observations. 
   In Fig.~\ref{rev_flux} the ISGRI fluxes in the 40--60 keV band are reported, averaged over 5 principal periods (rev. 48, rev. 49,
   rev. 93, rev. 104 and rev. 149). 
   A decrease by a factor of 1.8 has been observed also in JEM-X data in the 3--20 keV range within a period of 10 months.
   No JEM-X data are available in the intermediate period as the source was outside the FoV.  
   The $\chi^2$ test gives for the 40--60 keV band (20--40 keV) a $\chi^2 = 422$ ($\chi^2 = 1400$) with 4 d.o.f. which confirms the 
   high significance of the variability.
   A flux decrease has also been observed in the 2--12 keV range by \textit{RXTE}/ASM.
   The significance of the variability in these data is confirmed by a 
   $\chi^2 = 22$ (5 d.o.f) and thus a probability of $P_{\chi} = 5 \times 10^{-4}$, and
   the amplitude of the variation is about a factor of 10 with a very large
   uncertainty, due to the large error bars of these measurements.     
   \begin{figure}
   \centering
   {\includegraphics[angle=0,width=9.0cm]{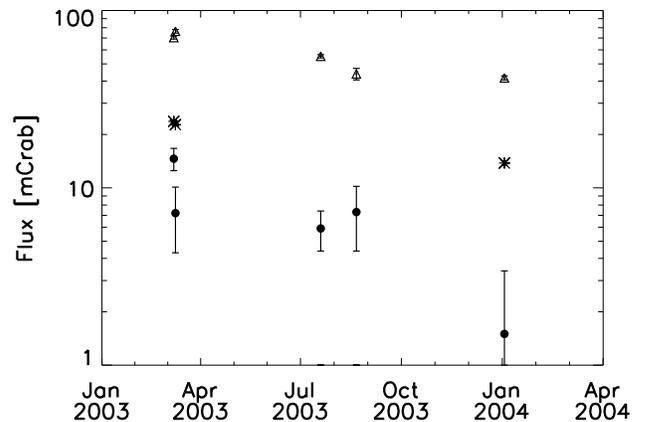}}
      \caption{Light curve of Centaurus~A during \textit{INTEGRAL} observations. Triangles are ISGRI
      fluxes in the 40--60 keV band 
      and stars are JEM-X fluxes in the 3--20 keV range. Circles are averaged \textit{RXTE}/ASM fluxes in 
      the 2--12 keV band during the days of the \textit{INTEGRAL} observations. The error bars for \textit{INTEGRAL}
      measurements are reported but smaller then the symbols used. 
      A decrease by a factor of 1.8 in the 3--20 keV and 40--60 keV bands was observed, while the 2--12 keV
      flux decreased by a factor of $\sim 10$.
       }
         \label{rev_flux}
   \end{figure} 


\section{Spectral Fit Results}\label{fit}

\subsection{\textit{INTEGRAL Spectra}}\label{INTEGRAL}
   
   The intrinsic high absorption column density ($>10^{22} \, \rm cm^{-2}$) typical for Seyfert 2 galaxies 
   has to be taken into account when fitting the spectra. 
   With the exception of Centaurus~A, for which a JEM-X spectrum has been extracted, the \textit{INTEGRAL} 
   source spectra do not cover energies below 20 keV, which makes it difficult to constrain the 
   intrinsic $N_{\rm H}$.
   Therefore during our analysis $N_{\rm H}$ was fixed to the values found in the literature, 
   i.e. $N_{\rm H} = 4 \times 10^{24} \rm \, cm^{-2}$ for
   {\mbox NGC 4945} and Circinus (\citealt{done03}, \citealt{matt99}), $2 \times 10^{23} \,\rm cm^{-2}$ 
   for {\mbox NGC 6300} \citep{matsumoto04} and $18 \times 10^{22} \, \rm cm^{-2}$ for {\mbox ESO 103--G35}
   \citep{wilkes01}. 
   The parameters of the best spectral fitting are reported in Table~\ref{INTEGRAL_fit}. 
   The spectra are shown in Figs~\ref{NGC_6300_spectrum}--\ref{CenA_149}, in photon units
   if only ISGRI data are avalaible and in count units when the SPI spectrum overlaps the ISGRI range. 
   
   A single power law corrected by photoelectric absorption (wabs model in XSPEC) is the best fit model for the \textit{INTEGRAL} 
   spectra of {\mbox NGC 6300}, {\mbox ESO 103--G35} and {\mbox NGC 4945} (Figs~\ref{NGC_6300_spectrum},
   \ref{ESO103_G35_spectrum} and \ref{NGC4945_spectrum}). 
   The photon indices are $\Gamma = 2.2 \pm 0.5$, $1.4 \pm 0.4$ and $1.9 \pm 0.1$, respectively. 
   
   The high signal-to-noise ratio of the spectrum of {\mbox NGC 4945} allows us to  
   study the presence of a high energy cut-off. The introduction of this component does not improve the quality of
   the fit ($\chi^2_{\nu}=1.4$ with 7 d.o.f. instead of $\chi^2_{\nu}=1.1$ with 8 d.o.f.) 
   and a lower limit of $E_c \sim 130 \, \rm keV$ can be given at $1 \,\sigma$ level.  
   If the $N_{\rm H}$ is let free to vary, it reaches a value of $~10^{25} \rm \, cm^{-2}$, with a slight 
   improvement of the $\chi^2_{\nu}$ which is not statistically significant (F-test probability of 0.5). 

   The combined ISGRI and SPI spectrum of Circinus (Fig.~\ref{Circinus_spectrum}) is best fitted with a high absorbed
   power law with $\Gamma = 1.8^{+0.4}_{-0.5}$ and a high energy exponential cut-off at $50^{+51}_{-18}$ keV
   which results in a $\chi^2_{\nu}$ of 1.1 (7 d.o.f.). This is significantly better than a simple power law fit 
   ($\chi^2_{\nu}=2.5$ with 8 d.o.f.). Introducing a Compton reflection component does not improve the fit
   results ($\chi^2_{\nu}=1.3$ with 6 d.o.f.). 
   
   The low significance of the detection of {\mbox MCG--05--23--16}
   hinders spectral extraction.
   In order to have an estimate of the flux level of this source, we assumed a spectrum identical to that of the Crab
   and we scaled the normalization using the count rate ratio,
   finding 13.5 and 15.6 mCrab (10.2 and 5.6 
   $\times 10^{-11} \rm \, erg \, cm^{-2} s^{-1}$) in the 20--40 keV and 40--60 keV bands, respectively. 
   \begin{figure}
   \centering
   {\includegraphics[angle=-90,width=8.0cm]{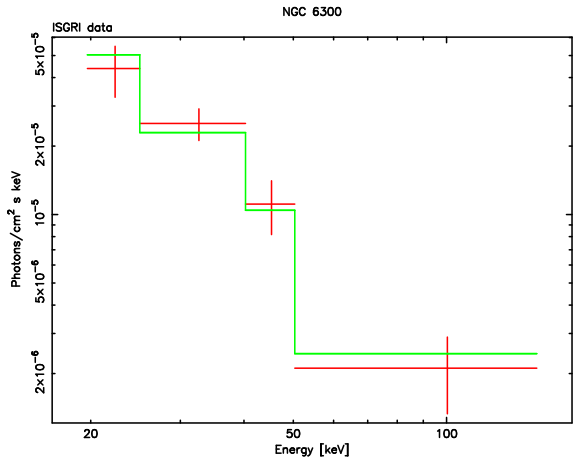}}
      \caption{\textit{INTEGRAL}/ISGRI spectrum of {\mbox NGC 6300}. This spectrum is best fitted by an absorbed power law with 
      $\Gamma = 2.2 \pm 0.5$.
              }
         \label{NGC_6300_spectrum}
   \end{figure}
   \begin{figure}
   \centering
   {\includegraphics[angle=-90,width=8.0cm]{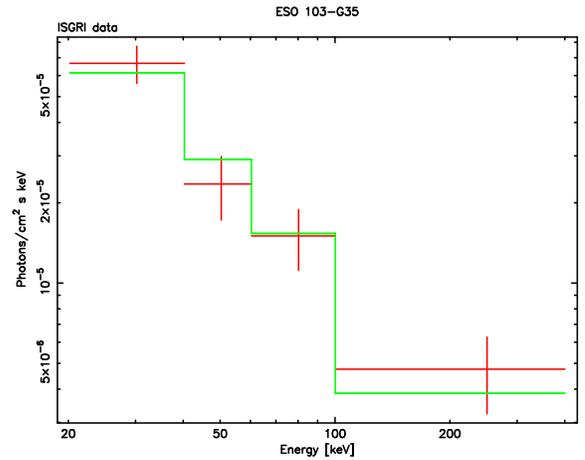}}
      \caption{\textit{INTEGRAL}/ISGRI spectrum of {\mbox ESO 103--G35}. An absorbed power law with $\Gamma = 1.4 \pm 0.4$ is the best 
      fit model for this spectrum.
              }
         \label{ESO103_G35_spectrum}
   \end{figure}
   \begin{figure}
   \centering
   {\includegraphics[angle=-90,width=8.5cm]{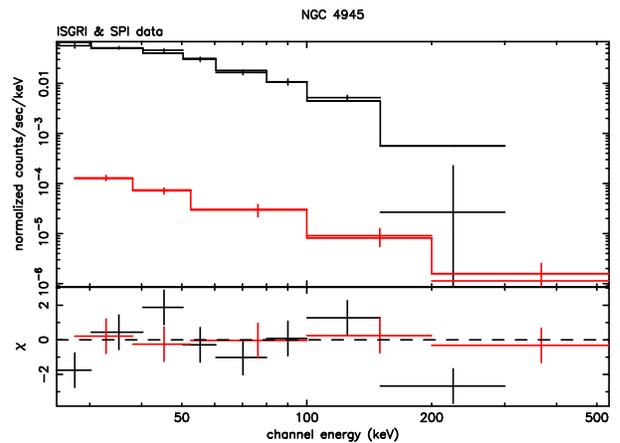}}
      \caption{ISGRI (upper points) and SPI (lower points) count spectra of {\mbox NGC 4945}. The data are well represented by an absorbed 
      power law with $\Gamma = 1.9 \pm 0.1$.
              }
         \label{NGC4945_spectrum}
   \end{figure}
   \begin{figure}
   \centering
   {\includegraphics[angle=-90,width=8.5cm]{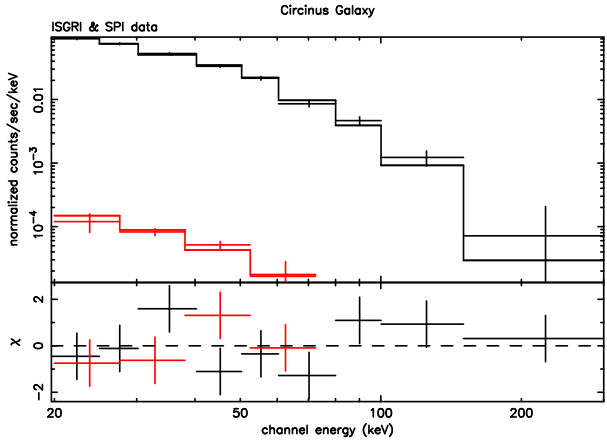}}
      \caption{ISGRI (upper points) and SPI (lower points) count spectra of the Circinus galaxy. The best fit model of this 
      \textit{INTEGRAL} spectrum is an absorbed power law with $\Gamma = 1.8^{+0.4}_{-0.5}$ and a high energy exponential 
      cut-off at $50^{+51}_{-18}$ keV.
              }
         \label{Circinus_spectrum}
   \end{figure}
\begin{table*}
\hspace{-0.7cm}
\begin{minipage}[t]{\columnwidth}
\caption{\textit{INTEGRAL} fit parameters}
\label{INTEGRAL_fit}
\renewcommand{\footnoterule}{}  
\centering 
\begin{tabular}{c c c c c c c c c} 
\hline\hline                
 Object & Instruments & Model & $\rm N_{\rm H}$ & $\Gamma$ & $\rm E_c$ & $\rm F_{20-100 \, keV}$\footnote{Observed fluxes from ISGRI data.} & $\rm F_{3-10 \, keV}$\footnote{Observed fluxes from JEM-X data.} & $\chi^2_{\nu} \, (dof)$ \\     
        &             &       & ($10^{22} \rm cm^{-2}$) &   &  (keV) & ($\rm 10^{-10} \, erg \, cm^{-2} s^{-1}$) & ($\rm 10^{-10} erg \, cm^{-2} s^{-1}$) &  \\    
\hline                        
MCG--05--23--16 & ISGRI          & --       & -- & -- & -- & $<$ 1.89 & -- & --     \\
NGC 4945        & ISGRI+SPI      & wabs po       & 400 & $1.9^{+0.1}_{-0.1}$ & $\gg 130$ & 2.2 & -- & 1.1(8)     \\
Centaurus A     & ISGRI+SPI+JEMX\footnote{High state.} & wabs po & $17^{+1}_{-1}$ & $2.02^{+0.03}_{-0.03}$ & -- & 8.9 & 3.0 & 1.2 (199) \\
Centaurus A     & ISGRI+SPI\footnote{Intermediate state.} & wabs po &  17            & $2.3^{+0.1}_{-0.1}$ & -- & 7.5 & --    & 3.3 (40)  \\
Centaurus A     & ISGRI+SPI+JEMX\footnote{Low state.} & wabs cutoffpl & $22^{+6}_{-6}$     & $1.8^{+0.2}_{-0.2} $ & $122^{+101}_{-39}$ & 4.9  & 1.4 & 1.3 (198) \\
Circinus galaxy & ISGRI+SPI      & wabs cutoffpl & 400 & $1.8^{+0.4}_{-0.5}$   & $50^{+51}_{-18}$ & 1.8 & $<$ 0.2 & 1.1(7) \\
NGC 6300        & ISGRI          & wabs po       & 22  & $2.2^{+0.5}_{-0.4}$   & --                & 0.5 & -- & 0.4(2)     \\
ESO 103--G35    & ISGRI          & wabs po       & 18  & $1.4^{+0.4}_{-0.4}$   & --                & 1.8 & -- & 0.7(2) \\
\hline                        
\end{tabular}
\end{minipage}
\end{table*}
\subsubsection{Centaurus A}\label{CenA}
   Unlike the other sources, Centaurus~A requires
   an analysis which takes into account flux variations over an interval of months.
   We split the data according to the flux level, choosing three main periods: high state (March 7--9, 2003,
   rev. 48--49), intermediate state (July 18 -- August 22, 2003, rev. 93 and 104) and low state (January 2--4, 2004, rev. 149). 
   We constructed a spectrum for each of these periods and fitted them separately (Figs~\ref{CenA_48_49}, \ref{CenA_104} and 
   \ref{CenA_149}). 
   
   The intermediate spectrum has been extracted mostly from data obtained during a staring observation (rev. 93).
   The problem with this strategy is that sometimes structures in the background
   can contribute to the overall count rate and can be mistaken for counts from the source. In dithering observations,
   this problem is minimized,   
   resulting in a time average of the background which decrease the probability to have repetitive structures.
   In fact a strong feature between 50 and 100 keV is present in the ISGRI spectrum of Centaurus~A in revolution 93,
   which has not been observed in the other spectra, nor in any other previous observation of this object.
   Therefore these data have been excluded from the intermediate state spectral analysis. 
   The source was outside the FoV of JEM-X, and no data were taken by SPI in this revolution. \\   
   Known features of the mask pattern in ISGRI are not well modelled by the response matrix and 
   could be important for sources as bright as Centaurus~A. 
   Instrumental background emission lines in the range 60--80 keV are known to be produced by Tungsten and Lead \citep{terrier03}.
   In order to avoid possible contamination of the Centaurus~A spectra, the energy bins in the 60--80 keV band 
   have been excluded from the analysis. 
   
   The combined JEM-X, ISGRI and SPI spectrum of the high state (Fig.~\ref{CenA_48_49}) 
   is best fitted by an absorbed power law with $N_{\rm H} = 17 \pm 1 \times 10^{22} \rm \, cm^{-2}$ and 
   $\Gamma = 2.02 \pm 0.03$. Adding a cut-off does not improve the fit 
   ($\chi^2_{\nu} = 1.2$, 198 d.o.f. instead of $\chi^2_{\nu} = 1.2$, 199 d.o.f.) and results in a 3$\sigma$
   lower limit at 270 keV for the cut-off energy. 
   
   The combined spectrum of the low state (Fig.~\ref{CenA_149}) is well represented ($\chi^2_{\nu} = 1.3$, 199 d.o.f.) by an absorbed power law 
   with significantly 
   higher absorption ($N_{\rm H} = (33 \pm 5) \times 10^{22} \rm \, cm^{-2}$) and photon index ($\Gamma = 2.17 \pm 0.06$) than in the high state.
   In Fig.~\ref{contour_CenA} we report the contour plots $\Gamma-N_{\rm H}$ for the high and low state of Centaurus~A, when a simple power law 
   is used to model both spectra. \\   
   A slight improvement ($\chi^2_{\nu} = 1.26$, 198 d.o.f., F-test probability of 0.002) is achieved when a cut-off is added to the fit model
   of the low state. This results in an absorption density $N_{\rm H} = (22 \pm 6) \times 10^{22} \rm \, cm^{-2}$, a photon index 
   $\Gamma = 1.8 \pm 0.2$ and a cut-off at $E_c = 122^{+100}_{-39}$ keV. Nevertheless, the energy of the cut-off and the absorption are not 
   well constrained with this model as it is evident in Fig.~\ref{contour_CenA2} where the $E_c - N_{\rm H}$ contour plots are shown. 
   The spectral differences between the high and low state are shown in the ratio between the two spectra in 
   Figure~\ref{CenA_ratio}. 
   The ratio in the range 5--80 keV is well approximated by a constant value of 1.7 ($\chi^2 = 7$, 7 d.o.f., $P_{\chi}=0.4$).
   A small variation is visible when we consider also the 3-5 keV bin ($P_{\chi}=0.002$ in the 5--80 keV), probably due
   to the change in the absorption.
   The analysis of the ratio in the 3--150 keV range shows a high significant variability with $\chi^2 = 160$ (10 d.o.f.), confirming
   a spectral change at high energies, consistent with the presence of a cut-off in the low state spectrum. 
   
   Adding a Gaussian feature to model a putative iron fluorescence line around 6.4 keV  
   does not improve the fit and results in a $3 \,\sigma$ upper limit for the line flux of $f_{\rm K\alpha} = 3.0$ and 
   $5.5 \times 10^{-3} \rm \, ph \, cm^{-2} s^{-1}$ in the high and low state, respectively. 
   
   After excluding revolution 93 from the analysis, only 23 ks of data are available for the intermediate 
   state of Centaurus~A. This ISGRI spectrum is therefore much less significant 
   ($40 \,\sigma$ in the 20--100 keV band) than the two in the high ($137 \,\sigma$) and low ($87 \, \sigma$) state which have 147 ks and
   172 ks of exposure time, respectively. In addition, during this observation Centaurus~A was always more than $> 9^{\circ}$ 
   from the centre of the ISGRI FoV. For these two reasons the combined ISGRI--SPI spectrum (no JEM-X data available)
   is best fitted with an absorbed power law 
   (with $N_{\rm H} = 17 \times 10^{22} \, \rm cm^{-2}$ fixed, which does not influence the fit result)
   with an unsatisfactory $\chi^2_{\nu}$ of 3.0 and a steep spectrum with a photon index of $2.3 \pm 0.1$ (Fig.~\ref{CenA_104}).
   No cut-off has been detected in the intermediate state. 
   
   A Compton reflection model has been applied to each of the three spectra of Centaurus~A. No improvement was found in the 
   $\chi^2$ of the fit and the reflection fraction was always below 0.1.    
   \begin{figure}
   \centering
   {\includegraphics[angle=-90,width=8.5cm]{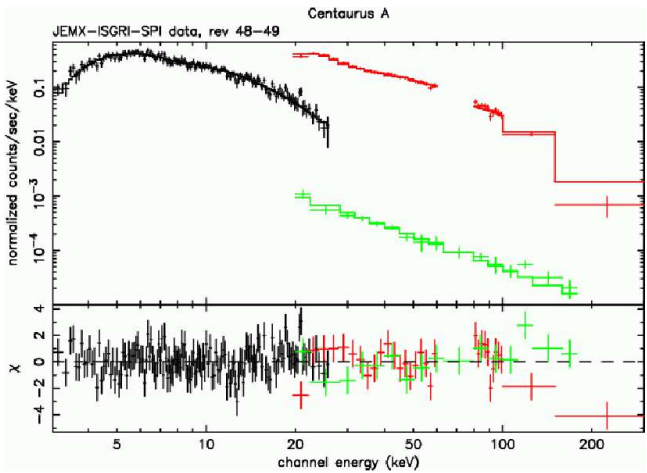}}
      \caption{Combined JEM-X, ISGRI (upper points above 20 keV) and SPI (lower points above 20 keV) spectrum of Centaurus~A during 
      the high state (revolutions 48--49). The best fit model is an absorbed power law with 
      $N_{\rm H} = 17 \pm 1 \times 10^{22} \rm \, cm^{-2}$ and $\Gamma = 2.02 \pm 0.03$.
              }
         \label{CenA_48_49}
   \end{figure}
   \begin{figure}
   \centering
   {\includegraphics[angle=-90,width=8.5cm]{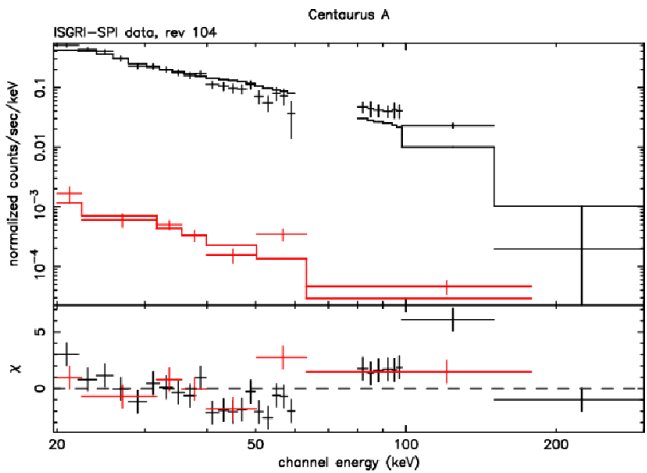}}
      \caption{ISGRI (upper points) and SPI (lower points) spectra of Centaurus~A during the intermediate state (revolution 104). 
      As the source was at $> 9^{\circ}$ from the centre of the FoV, no JEM-X data are available for this data set and the best fit model
      (an absorbed power law with $\Gamma = 2.3 \pm 0.1$) provides an unsatisfactory $\chi^2_{\nu}$ of 3.0. 
              }
         \label{CenA_104}
   \end{figure}
   \begin{figure}
   \centering
   {\includegraphics[angle=-90,width=8.8cm]{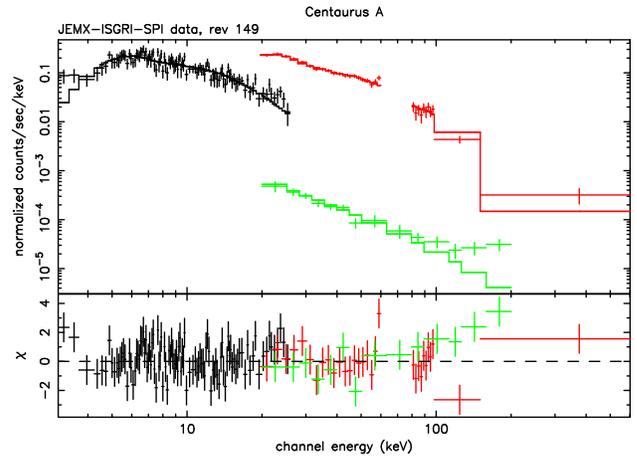}}
      \caption{Combined JEM-X, ISGRI (upper points above 20 keV) and SPI (lower points above 20 keV) spectrum of Centaurus~A 
      during the low state (revolution 149). These \textit{INTEGRAL} data can be modelled equally well by an absorbed power law with cut-off
      at $E_c = 122^{+100}_{-39}$ keV (model reported in the figure) or by a simple absorbed power law with 
      $N_{\rm H} = (33 \pm 5) \times 10^{22} \rm \, cm^{-2}$, a factor 2 higher than during the high state.
              }
         \label{CenA_149}
   \end{figure}
   \begin{figure}
   \centering
   {\includegraphics[angle=-90,width=8.0cm]{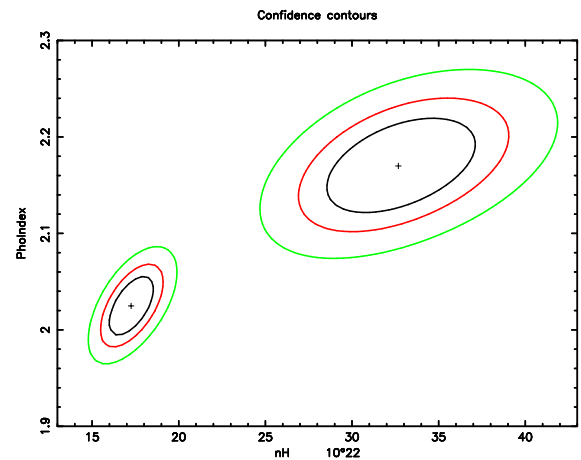}}
      \caption{$\Gamma$--$N_{\rm H}$ contour plots at 68\%, 90\% and 99\% confidence level for the \textit{INTEGRAL} spectra of 
      Centaurus~A in the high (curves on the left) and low state (right).
              } 
         \label{contour_CenA}
   \end{figure}
   \begin{figure}
   \centering
   {\includegraphics[angle=-90,width=8.0cm]{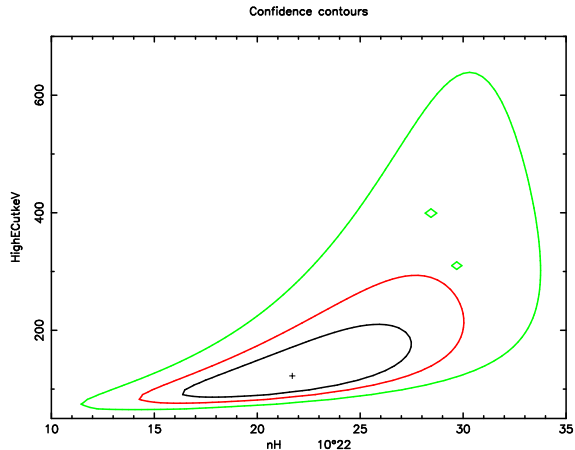}}
      \caption{$E_c$--$N_{\rm H}$ contour plots at 68\%, 90\% and 99\% confidence level for the \textit{INTEGRAL} spectra of 
      Centaurus~A in the low state.
              } 
         \label{contour_CenA2}
   \end{figure}
   \begin{figure}
   \centering
   {\includegraphics[angle=0,width=8.5cm]{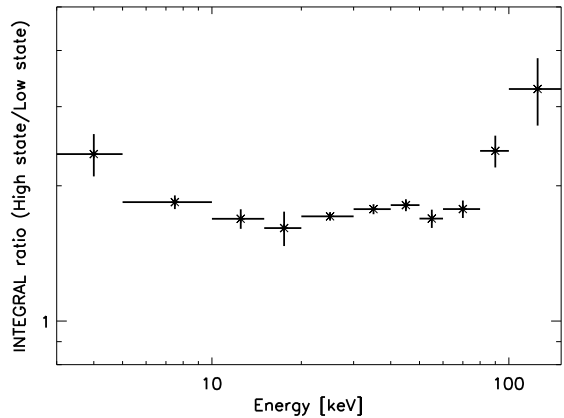}}
      \caption{Count rate ratio for Centaurus~A between \textit{INTEGRAL} data extracted during the high
      and the low state. JEM-X and ISGRI spectra have been used.
              }
         \label{CenA_ratio}
   \end{figure}
\subsubsection{$\gamma$-ray Annihilation Line}
   The SPI spectrometer offers the opportunity to look for line features. In case the high energy 
   spectrum of AGN would be dominated by non-thermal processes, one could expect a significant amount 
   of $\rm e^\pm$ pair annihilation processes around $511 \rm \, keV$. At this energy, SPI has a 
   resolution of $FWHM = 1.95 \rm \, keV$ \citep{attie03}. 
   Imaging analysis in a 50 keV wide 
   bin around the restframe energy of the annihilation line has been performed on the three brightest sources of our sample. 
   It has been possible to obtain only a 3$\sigma$ upper limit for the flux in the 485--535 keV band of 
   4.5, 4.4 and 3.1 $\times 10^{-4} \, \rm ph \, cm^{-2} \, s^{-1}$ 
   for Centaurus~A, {\mbox NGC 4945} and the Circinus galaxy, respectively.
%
\subsection{Comparison with \textit{BeppoSAX}/PDS Results}\label{SAX}
    As most of the data we discussed have been taken with ISGRI in the hard X-ray range above 20 keV, the high 
    energy instrument of \textit{BeppoSAX}, PDS, has been used to make a comparison in the same energy band.
    It should be taken into account that the PDS spectra alone cannot well constrain components like absorption or 
    Compton reflection. For the PDS analysis we fixed $N_{\rm H}$ at the value found in the 
    literature for {\mbox MCG--05--23--16} ($N_{\rm H} = 3 \times 10^{22} \, \rm cm^{-2}$, \citealt{mattson04}).
    For {\mbox NGC 4945}, Centaurus~A and Circinus the values we found for the PDS spectra are in agreement with what 
    has been reported for these sources in the \textit{BeppoSAX} broad band analysis. The cases of {\mbox NGC 6300} and {\mbox ESO 103--G35} are 
    discussed further in this section.  \\    
    For \textit{BeppoSAX} broad band analyses of the sources in the sample,    
    which are beyond the scope of this work, we refer to specific papers,
    i.e. \citet{mattson04} for {\mbox MCG--05--23--16}, \citet{guainazzi00} for {\mbox NGC 4945}, \citet{grandi03} for Centaurus~A,
    \citet{matt99} and \citet{guainazzi99} for the Circinus galaxy, \citet{guainazzi02} for {\mbox NGC 6300} and \citet{wilkes01} 
    for {\mbox ESO 103--G35}. \\
    The best fit parameters of the PDS spectra are reported in Table~\ref{SAX_tab}.
\subsubsection{The Circinus galaxy}      
    The Circinus galaxy reveals also in the PDS data the presence of a high energy cut-off at $36^{+21}_{-10}$ keV, compatible with
    the ISGRI value of $50^{+51}_{-18}$ keV. The $E_c$--$\Gamma$ contour plots obtained from \textit{INTEGRAL} (solid line) and PDS
    data (dotted line) are reported in Figure~\ref{contour}. Both the \textit{INTEGRAL} and PDS data sets are consistent with a high energy cut-off
    in the range 35--55 keV at 99\% confidence level. 
    We performed a fit of \textit{INTEGRAL} and PDS spectra also with the reflection model used by \citet{matt99} to fit the combined spectrum 
    of all \textit{BeppoSAX} instruments. Fixing all the parameters but the normalization at the value reported by \citet{matt99}, 
    the fit results in a worse $\chi^2_{\nu}$ (2.5 with 9 d.o.f for the \textit{INTEGRAL} spectrum and 11.4 with 7 d.o.f. for the PDS one) compared
    to a simple cut-off power law which confirms the difficulty in studying the reflection component modelling only the high energy data. 
    Nevertheless the value of the photon index and the cut-off energy in the model of \citet{matt99} 
    are consistent with the values we find for these parameters by fitting the \textit{INTEGRAL} and PDS spectra only.
\subsubsection{{\mbox NGC 6300}}               
    As discussed in \citealt{guainazzi02}, the shape of the \textit{BeppoSAX} spectrum of {\mbox NGC 6300} can be explained by the presence of a 
    Compton-thick absorber with a high $N_{\rm H}$ of about $3 \times 10^{24} \, \rm cm^{-2}$ or by a Compton-thin absorber 
    ($N_{\rm H} \sim 3 \times 10^{23} \, \rm cm^{-2}$) added to a Compton reflection component with a high reflection fraction ($\sim 4.2$).
    The first model applies better to our PDS data in the 15--150 keV region with $N_{\rm H} = 472^{+142}_{-130} \times 10^{22} \, \rm cm^{-2}$ 
    and the photon index found is consistent with the one obtained from ISGRI data (even when in the fitting model of the ISGRI spectrum 
    the absorption is fixed to the PDS value). 
    The difference by a factor of $\sim 20$ between the $N_{\rm H}$ used in the ISGRI data fit (best value reported by \citealt{guainazzi02})
    and the one found in PDS data cannot be considered significant due to the lack of data below 20 keV in our analysis.
    The value of the photon index and the lack of a cut-off below 200 keV observed in ISGRI and PDS spectra are characteristics common also
    to the best fit model used by \citet{guainazzi02} for the \textit{BeppoSAX} spectrum. Fitting PDS data with this model, which includes a 
    reflection
    component, results in an unacceptable fit: $\chi^2_{\nu}$ = 7 (9 d.o.f). $\chi^2_{\nu}$ = 0.06 (3 d.o.f) is obtained for ISGRI data,
    but the low statistics of this spectrum does not allow us to draw a firm conclusion about this model. 
\subsubsection{Centaurus~A}           
    The best fit model for the PDS spectrum of Centaurus~A is an absorbed power law, where the value of the photon index is in agreement
    with what was found by \textit{INTEGRAL} during the low state, when the source was at the same flux level as 
    in the \textit{BeppoSAX} data. The $N_{\rm H}$ is not well constrained and is compatible with the one obtained from 
    \textit{INTEGRAL} spectra. Adding a cut-off does not improve the $\chi^2$ and gives values of $E_c \gg$ 300 keV. 
    The results from the broad band analysis of \textit{BeppoSAX} data \citep{grandi03} are in agreement with what found by the PDS alone.
    No high energy cut-off nor evidence for Compton reflection has been found in this \textit{BeppoSAX} analysis. 
\subsubsection{{\mbox MCG--05--23--16}}        
    For {\mbox MCG--05--23--16} only a comparison of the flux level is possible, and this shows that \textit{BeppoSAX} caught the source at about 
    the same level as \textit{INTEGRAL} did, with a flux of 6.7 and 4.0 $\times 10^{-11} \rm \, erg \, cm^{-2} s^{-1}$ in the 20--40 and 40--60 keV 
    bands, respectively. 
    The PDS spectrum alone is best fitted by an absorbed power law with photon index $\Gamma = 1.4 \pm 0.3$ and cut-off at $E_c = 60^{+56}_{-21}$ 
    keV.
    Fixing the parameters, the reflection model of \citet{mattson04} for the \textit{BeppoSAX} data represents the PDS data as well 
    ($\chi^2_{\nu}$ = 1.2 with 8 d.o.f), but the spectrum is softer ($\sim$ 1.7) and the cut-off appears at higher energies (fixed at 100 keV by 
    \citealt{mattson04}). 
\subsubsection{{\mbox ESO 103--G35}}               
    The PDS spectrum of {\mbox ESO 103--G35} is well modeled by an absorbed power law with high energy cut-off ($\chi^2_{\nu} = 0.3$, 4 d.o.f.), 
    but the values of $\Gamma$ and $E_c$ are not constrained, spanning at 3$\sigma$ level the ranges 0.5--2 and 20--180 keV, respectively.
    Therefore, we decided to fit the spectrum with the best fit model of the combined \textit{BeppoSAX} data \citep{wilkes01},
    even if, as reveled by our analysis of the PDS data, the spectrum reported by \citet{wilkes01} should be contaminated by the presence of 
    a source in the +OFF background field.
    The model used requires a 
    photon index of 1.7, a cut-off with energy of $E_c = 29$ keV and $e$--folding energy of 40 keV (we freely varied only the normalisation).
    The signal-to-noise ratio of the ISGRI spectrum of {\mbox ESO 103--G35} is not high enough to investigate the
    presence of a high energy cut-off, and the ISGRI photon index is compatible with the \textit{BeppoSAX} one. 
\subsubsection{{\mbox NGC 4945}}              
    The PDS spectrum of {\mbox NGC 4945} is best modeled by a highly ($N_{\rm H} = 5 \times 10^{24} \, \rm cm^{-2}$) absorbed power law 
    with a photon index $\Gamma = 1.6 \pm 0.3$ and a not well constrained high energy cut-off at $E_c = 129^{+252}_{-55}$ keV. 
    The lower limit obtained from \textit{INTEGRAL} data (130 keV) for the cut-off energy is consistent with what found by the PDS and the
    photon index is steeper ($\Gamma = 1.9 \pm 0.3$) but still compatible at 3$\sigma$ with the PDS one. 
    These values are in agreement with the results obtained by \citet{guainazzi00} for the broad band \textit{BeppoSAX} spectrum
    ($N_{\rm H} = 4.5\pm 0.7 \times 10^{24} \, \rm cm^{-2}$, $\Gamma = 1.59^{+0.17}_{-0.37}$, $E_c = 140^{+100}_{-50}$ keV).

\begin{table*}
\begin{minipage}[t]{\columnwidth}
\caption{\textit{BeppoSAX}/PDS fit parameters}
\label{SAX_tab}
\renewcommand{\footnoterule}{}  
\centering 
\begin{tabular}{c c c c c c c} 
   \hline\hline                 
 Object & Model & $N_{\rm H}$ & $\Gamma$   &  $E_c$ & $F_{20-100 \, \rm keV}$\footnote{Observed fluxes.} & $\chi^2_{\nu} (dof)$ \\    
        &    & ($10^{22} \rm cm^{-2}$) &   &  (keV) & ($\rm 10^{-10} \, erg \, cm^{-2} s^{-1}$) &  \\    
   \hline                        
MCG--05--23--16 & wabs cutoffpl & 3		      & $1.4^{+0.3}_{-0.3}$    & $60^{+56}_{-21}$   & 1.5 & 0.6 (6) \\	       
NGC 4945        & wabs cutoffpl & $500^{+115}_{-118}$ & $1.6^{+0.3}_{-0.3}$    & $129^{+252}_{-55}$ & 2.9 & 1.2 (5) \\	       
Centaurus A     & wabs po	& $32^{+17}_{-17}$    & $1.89^{+0.03}_{-0.03}$ & --		    & 5.3 & 1.4 (7) \\	       
Circinus galaxy & wabs cutoffpl & $519^{+122}_{-110}$ & $1.5^{+0.5}_{-0.5}$    & $36^{+21}_{-10}$   & 1.9 & 0.3 (5) \\	       
NGC 6300        & wabs po	& $472^{+142}_{-130}$ & $2.2^{+0.2}_{-0.2}$    & --		    & 1.3 & 0.8 (7) \\	       
ESO 103--G35    & wabs po highe &  18		        & 1.74                 & 29\footnote{$E_{fold} = 40$ keV.} & 0.7 & 0.5 (6) \\	       
   \hline                                   
\end{tabular}
\end{minipage}
\end{table*}
   \begin{figure}
   \centering
   {\includegraphics[angle=0,width=8.0cm]{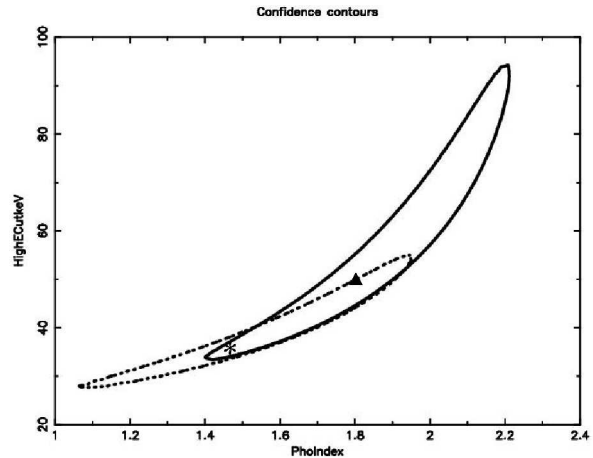}}
      \caption{$E_c$--$\Gamma$ contour plots at 99\% confidence level for \textit{INTEGRAL} (solid line for the contour and triangle for 
      the values of the best fit) and \textit{BeppoSAX} (dotted line and star) data of the Circinus galaxy.
              } 
         \label{contour}
   \end{figure}


\section{Discussion on individual sources.}\label{discussion}
   The \textit{BeppoSAX} observation was the first one above 10 keV for the Circinus galaxy and shows
   a strong absorption with $N_{\rm H} \sim 4-5 \times 10^{24} \rm cm^{-2}$. 
   \textit{INTEGRAL} data do not allow to constrain the value of $N_{\rm H}$, but adding this component
   fixed at $N_{\rm H} = 4 \times 10^{24} \, \rm cm^{-2}$ improves the fit significantly 
   ($\chi^2_{\nu}$ from 1.8 to 1.1). 
   Such a high absorption can originate in the molecular
   torus surrounding the nucleus or in the hot plasma which scatters the optical broad lines, and
   allows the observation of the direct emission only above 10 keV.
   A high energy cut-off at $E_c \sim$ 35--55 keV is required in both \textit{INTEGRAL} and \textit{BeppoSAX} data
   (\citealt{matt99} and this work).  
   This results in a temperature of $T \sim 4-6 \times 10^8 \rm \, K$ for the distribution of thermal 
   electrons in the Comptonizing medium. The presence of a cut-off supports the scenario in which the 
   X-ray emission of Seyfert galaxies originates in a two-phases accretion disk, where the soft photons emitted by a cold
   ($kT \la 50 \, \rm eV$) optically thick disk are Comptonized in a hot ($kT = 50-500 \, \rm keV$) region, for example an 
   optically thin corona above the disk \citep{haardt93}. 
   A Compton reflection component observed in the broad band \textit{BeppoSAX} data (\citealt{matt99} and \citealt{guainazzi99})
   has not been detected by \textit{INTEGRAL} because of the lack of data below 20 keV. \\   
   
   Several observations of NGC 4945 have been performed in the hard X-ray band. 
   \citet{Done96} presented a combined study of \textit{ASCA}, \textit{Ginga} and OSSE data in which they found    
   a continuum with a photon index $\Gamma = 1.9$ and a lower limit
   for the cut-off energy at $E_c = 270 \,\rm keV$.  
   A high energy cut-off in the range 100--300 keV has been detected in \textit{BeppoSAX} and \textit{RXTE} data
   (\citealt{guainazzi00}, \citealt{madejski00}).
   The \textit{INTEGRAL} spectrum is well fitted with a highly
   absorbed power law, and adding a cut-off component to the model does not improve the fit and results in a lower limit 
   of $E_c \sim 130 \,\rm keV$, consistent with what has been found by previous missions. 
   A preliminary analysis of a 250 ks \textit{INTEGRAL} observation in January 2004 (revolution 149 and 150)
   confirms the lack of a cut-off below 150 keV. \\
   Hard X-ray variability on time scales of a day, observed in \textit{BeppoSAX} and \textit{RXTE} data, 
   is confirmed by a 20\% flux decrease within 2 days detected by \textit{INTEGRAL} in March 2003. \\
      
   NGC 6300 is the weakest source in our sample with a flux $F(20-100 \, \rm keV) = 5.5 \times 10^{-11} \rm \, erg \, cm^{-2} \, s^{-1}$,
   2 times lower than during \textit{BeppoSAX} observations in 1999.
   This low state can be expected in the frame proposed by \citealt{guainazzi02} in which NGC 6300 
   represents a case of a transient AGN, undergoing two
   periods of low and high activity as seen also in other AGN (e.g. NGC 4051 and Mkn 3). 
   The double nature of the Compton-thin and Compton-thick object (\citealt{guainazzi02}, \citealt{leighly99})
   cannot be studied with the \textit{INTEGRAL} data available to date. 
   Longer exposure time and a spectrum which extends also below 20 keV are necessary to study these behaviours. 
   In agreement with the results found in \textit{BeppoSAX} data (\citealt{guainazzi02} and this work), 
   the ISGRI spectrum does not reveal the presence of a high energy cut-off up to 150 keV. \\
              
   Centaurus~A is peculiar among AGN as it has been classified as a low luminosity 
   Fanaroff-Riley I galaxy \citep{fanaroff74}, a misdirected BL Lac object \citep{morganti92} and a 
   Seyfert 2 object \citep{dermer95}.
   The spectrum of Centaurus~A in the X-ray band is well represented by a heavily absorbed power law continuum
   with an iron line at 6.4 keV, features typical of Seyfert 2 spectra. 
   In contrast to other Seyfert 2 galaxies, the iron line is not associated with a strong Compton reflection continuum 
   (\citealt{rothschild99}, \citealt{grandi03}) and it is thought to be produced in a cold and optically thin 
   circumnuclear material \citep{benlloch01}. 
   \textit{INTEGRAL} observations confirm the lack of a Compton reflection, as adding this component does not improve the fits and results
   in a reflection fraction $<$0.1. 
   Only an upper limit can be derived from JEM-X data for the fluorescence iron line at 6.4 keV, with a flux an order of  
   magnitude above the values found in the literature (\citealt{grandi03}, \citealt{rothschild99}). \\
   The absorption has been found to assume values from $N_{\rm H} = 8 \times 10^{23} \rm \, cm^{-2}$ \citep{benlloch01}
   to $N_{\rm H} = 35 \times 10^{23} \rm \, cm^{-2}$ \citep{turner97}, range in which also \textit{INTEGRAL} results are placed. \\
   Centaurus~A is known to be a highly variable object on both long and short time scales \citep{jourdain93}. 
   Within 10 months, the hard X-ray flux of Centaurus~A has changed by a factor of $\sim 2$ in \textit{INTEGRAL} data.
   The flux variation is associated with spectral variation,    
   which can be modelled equally well by an increase by a factor of
   2 of the absorption or by the presence of a high energy cut-off at $E_c \ga$ 120 keV in the low state spectrum.
   Variations of the absorption are a common characteristic in
   Seyfert 2 galaxies, with changes of 20--80\% on a one year time scales \citep{risaliti02b}. Small variations of $N_{\rm H}$
   ($<$ 30\%) have been already observed for Centaurus~A in \textit{BeppoSAX} \citep{grandi03} and \textit{RXTE} data \citep{benlloch01}. 
   It should be also taken into account that    
   the flux decrease of an unknown or unresolved component in the soft X-ray band can mimick 
   an increase of the absorption \citep{grandi03}. \\
   The ratio between the high and low state spectra shows an increase at energies above 80 keV consistent with the presence of a cut-off at 122 keV,
   which is however an unusual feature for Centaurus~A. 
   Hints for a break or a cut-off in the hard X-ray and soft $\gamma$-ray spectra have been found by OSSE, \textit{BeppoSAX}, and 
   \textit{RXTE} data (\citealt{steinle98}, \citealt{grandi03}, \citealt{benlloch01}), 
   but these studies place this feature in the 300--700 keV range. 
   The high energy cut-off combined with the other characteristics of this object (iron line feature, lack of reflection,
   lack of correlation between continuum and iron line variability) suggests the presence of a hot, thick and optically thin 
   accretion flow in the nucleus of Centaurus~A \citep{grandi03}.\\     

\section{Conclusions}\label{conclusions}
   The spectral characteristics of our \textit{INTEGRAL} sample can be generally summarized as follows: 
   a hard X-ray continuum emission described by a power law
   with a wide range of photon indices ($\Gamma \sim 1.4-2.3$) and, in the case of Circinus 
   and Centaurus~A the presence of a high energy cut-off.
   The average photon index $\Gamma = 1.9 \pm 0.3$ obtained from \textit{INTEGRAL} spectra is consistent 
   with the values found for other samples of Seyfert 2 galaxies in \textit{BeppoSAX} 
   (\citealt{risaliti02}, \citealt{deluit03}, \citealt{malizia03} and this work),      
   OSSE \citep{zdziarski00} and \textit{Ginga} \citep{smith96} data. \\
   Only in certain cases adding a high energy cut-off to the fit model improves the results for average
   spectra (\citealt{zdziarski00}, \citealt{malizia03}), while in others no cut-off is required, suggesting 
   a wide range for the energy of this feature \citep{deluit03}. 
   Studies of single objects confirm that
   the cut-off in the 100--300 keV range is not a universal characteristic of all Seyfert 2 \citep{risaliti02}, 
   and the results for our sample support these findings.
   \textit{INTEGRAL} confirms the presence of a cut-off at $\sim$50 keV for the Circinus galaxy, a lower limit of 130 keV for
   NGC 4945 and the lack of this feature for NGC 6300, in agreement with what has been found in PDS spectra.
   A poorly constrained cut-off at $\ga$ 120 keV has been detected for Centaurus~A during the \textit{INTEGRAL} low state,
   but this feature has not been seen in the other \textit{INTEGRAL} observations reported in this work and in the PDS spectrum.
   Cut-offs below 100 keV have been found for {\mbox MCG--05--23--16} and {\mbox ESO 103--G35} by PDS, but could not be studied
   by \textit{INTEGRAL} because of the short exposure time of those observations. 
      
   The cut-off energy in AGN is an important parameter for the models of the cosmic X-ray background (XRB). 
   Consistent results have been achieved for example by the models of \citealt{gilli99} and \citealt{treister05}, in which a high energy cut-off at
   300 keV has been assumed.    
   The existence of AGN with a cut-off at significantly lower energies should be taken into account in order to more accurately estimate
   the contribution of the AGN to the XRB at energies $E \ga 80 \rm \, keV$ where
   the choice of this parameter is more relevant \citep{treister05}.
   Examples for these AGN are the Circinus galaxy, {\mbox ESO 103--G35} and {\mbox MCG--05--23--16} with cut-off energies in the
   30--60 keV range, found by \textit{INTEGRAL} and \textit{BeppoSAX}.
   It would be important to investigate if this type of objects constitutes only extreme cases or the 
   prototypes of a population with an cut-off energy below 100 keV. \\
   
   In spite of the fact that the power law and the cut-off power law are phenomenological models,
   the detection of a cut-off provides physical information about the presence of the Comptonization in a source.
   Moreover, the presence of a cut-off in Seyfert 2 galaxies associated with the observation of Polarized Broad Lines (PBL)
   could be an indication that these objects have a "Seyfert 1 nucleus" and then    
   a hidden Broad Line Region, whereas the objects without PBL 
   could have nuclear properties intrinsically different from those of Seyfert 1 (\citealt{antonucci85}, \citealt{tran95}).
   In fact a comparison between two samples of these Seyfert 2 subclasses (with and without PBL) and a sample of Seyfert 1 
   show that Seyfert 2 with PBL present 
   characteristics similar to those of Seyfert 1, i.e. the presence of a cut-off and the hardness of the spectrum \citep{deluit04}. \\         
   More meaningful physical models, like Comptonization or reflection are too complex to be adequately constrained by the low 
   signal-to-noise data we have up to now.
          
   \textit{INTEGRAL} has already detected 38 AGN (up to January 2004, \citealt{beckmann05b}), among which there are 15 Seyfert 2 galaxies
   and further \textit{INTEGRAL} observations, both pointed and during regular scans of the Galactic Plane, will allow to investigate deeper
   the primary nuclear emission of Seyfert 2, as already done for a number of AGN 
   (\citealt{courvoisier03b}, \citealt{beckmann04}, \citealt{pian05}, \citealt{derosa05}, \citealt{beckmann05a}),
   and to compare their characteristics with those of Seyfert 1 objects. \\ 
     

\begin{acknowledgements}
   We thank L. Piro to have encouraged the development of this work
   and C. Shrader for proof-reading of the manuscript. We would like 
   to thank also the referee E. Pian for the valuable suggestions 
   which helped us to improve the paper.
\end{acknowledgements}


\bibliographystyle{aa}
\bibliography{biblio}

\begin{thebibliography}{64}
\expandafter\ifx\csname natexlab\endcsname\relax\def\natexlab#1{#1}\fi

\bibitem[{{Antonucci} \& {Miller}(1985)}]{antonucci85}
{Antonucci}, R.~R.~J. \& {Miller}, J.~S. 1985, \apj, 297, 621

\bibitem[{{Arnaud}(1996)}]{arnaud96}
{Arnaud}, K.~A. 1996, in ASP Conf. Ser. 101: Astronomical Data Analysis
  Software and Systems V, 17

\bibitem[{{Atti{\' e}} {et~al.}(2003){Atti{\' e}}, {Cordier}, {Gros},
  {Laurent}, {Schanne}, {Tauzin}, {von Ballmoos}, {Bouchet}, {Jean}, {Kn{\"
  o}dlseder}, {Mandrou}, {Paul}, {Roques}, {Skinner}, {Vedrenne}, {Georgii},
  {von Kienlin}, {Lichti}, {Sch{\" o}nfelder}, {Strong}, {Wunderer}, {Shrader},
  {Sturner}, {Teegarden}, {Weidenspointner}, {Kiener}, {Porquet}, {Tatischeff},
  {Crespin}, {Joly}, {Andr{\' e}}, {Sanchez}, \& {Leleux}}]{attie03}
{Atti{\' e}}, D., {Cordier}, B., {Gros}, M., {et~al.} 2003, \aap, 411, L71

\bibitem[{{Bassani} {et~al.}(2004){Bassani}, {Malizia}, {Stephen}, {Gianotti},
  {Schiavone}, {Bazzano}, {Bird}, {Bouchet}, {Courvoisier}, {Dean}, {De
  Cesare}, {Del Santo}, {De Rosa}, {Hudec}, {Mirabel}, {Laurent}, {Piro},
  {Shaw}, \& {Zdziarski}}]{bassani04}
{Bassani}, L., {Malizia}, A., {Stephen}, J., {et~al.} 2004, Proceedings of the
  5th INTEGRAL Workshop, ESA SP-552, p.139, [astro-ph/0404442]

\bibitem[{{Beckmann} {et~al.}(2004){Beckmann}, {Gehrels}, {Favre}, {Walter},
  {Courvoisier}, {Petrucci}, \& {Malzac}}]{beckmann04}
{Beckmann}, V., {Gehrels}, N., {Favre}, P., {et~al.} 2004, \apj, 614, 641

\bibitem[{{Beckmann} {et~al.}(2005{\natexlab{a}}){Beckmann}, {Gehrels},
  {Shrader}, {Soldi}, P., {Zdziarski}, {Petrucci}, \& {Malzac}}]{beckmann05a}
{Beckmann}, V., {Gehrels}, N., {Shrader}, C.~R., {et~al.} 2005{\natexlab{a}},
  accepted by ApJ, [astro-ph/0508327]

\bibitem[{{Beckmann} {et~al.}(2005{\natexlab{b}}){Beckmann}, {Shrader},
  {Gehrels}, \& {Soldi}}]{beckmann05b}
{Beckmann}, V., {Shrader}, C.~R., {Gehrels}, N., \& {Soldi}, S.
  2005{\natexlab{b}}, submitted to ApJ

\bibitem[{{Benlloch} {et~al.}(2001){Benlloch}, {Rothschild}, {Wilms},
  {Reynolds}, {Heindl}, \& {Staubert}}]{benlloch01}
{Benlloch}, S., {Rothschild}, R.~E., {Wilms}, J., {et~al.} 2001, \aap, 371, 858

\bibitem[{{Bird} {et~al.}(2004){Bird}, {Barlow}, {Bassani}, {Bazzano},
  {Bodaghee}, {Capitanio}, {Cocchi}, {Del Santo}, {Dean}, {Hill}, {Lebrun},
  {Malaguti}, {Malizia}, {Much}, {Shaw}, {Stephen}, {Terrier}, {Ubertini}, \&
  {Walter}}]{bird04}
{Bird}, A.~J., {Barlow}, E.~J., {Bassani}, L., {et~al.} 2004, \apjl, 607, L33

\bibitem[{{Chiappetti} \& {dal Fiume}(1997)}]{chiappetti97}
{Chiappetti}, L. \& {dal Fiume}, D. 1997, in Data Analysis in Astronomy IV, 101

\bibitem[{{Courvoisier} {et~al.}(2003{\natexlab{a}}){Courvoisier}, {Beckmann},
  {Bourban}, {Chenevez}, {Chernyakova}, {Deluit}, {Favre}, {Grindlay}, {Lund},
  {O'Brien}, {Page}, {Produit}, {T{\" u}rler}, {Turner}, {Staubert},
  {Stuhlinger}, {Walter}, \& {Zdziarski}}]{courvoisier03b}
{Courvoisier}, T.~J.-L., {Beckmann}, V., {Bourban}, G., {et~al.}
  2003{\natexlab{a}}, \aap, 411, L343

\bibitem[{{Courvoisier} {et~al.}(2003{\natexlab{b}}){Courvoisier}, {Walter},
  {Beckmann}, {Dean}, {Dubath}, {Hudec}, {Kretschmar}, {Mereghetti},
  {Montmerle}, {Mowlavi}, {Paltani}, {Preite Martinez}, {Produit}, {Staubert},
  {Strong}, {Swings}, {Westergaard}, {White}, {Winkler}, \&
  {Zdziarski}}]{courvoisier03a}
{Courvoisier}, T.~J.-L., {Walter}, R., {Beckmann}, V., {et~al.}
  2003{\natexlab{b}}, \aap, 411, L53

\bibitem[{{De Rosa} {et~al.}(2005){De Rosa}, {Piro}, {Tramacere}, {Massaro},
  {Walter}, {Bassani}, {Malizia}, {Bird}, \& {Dean}}]{derosa05}
{De Rosa}, A., {Piro}, L., {Tramacere}, A., {et~al.} 2005, \aap, 438, 121

\bibitem[{{Deluit} \& {Courvoisier}(2003)}]{deluit03}
{Deluit}, S. \& {Courvoisier}, T.~J.-L. 2003, \aap, 399, 77

\bibitem[{{Deluit}(2004)}]{deluit04}
{Deluit}, S.~J. 2004, \aap, 415, 39

\bibitem[{{Dermer} \& {Gehrels}(1995)}]{dermer95}
{Dermer}, C.~D. \& {Gehrels}, N. 1995, \apj, 447, 103

\bibitem[{{Di Cocco} {et~al.}(2003){Di Cocco}, {Caroli}, {Celesti}, {Foschini},
  {Gianotti}, {Labanti}, {Malaguti}, {Mauri}, {Rossi}, {Schiavone},
  {Spizzichino}, {Stephen}, {Traci}, \& {Trifoglio}}]{dicocco03}
{Di Cocco}, G., {Caroli}, E., {Celesti}, E., {et~al.} 2003, \aap, 411, L189

\bibitem[{{Diehl} {et~al.}(2003){Diehl}, {Baby}, {Beckmann}, {Connell},
  {Dubath}, {Jean}, {Kn{\" o}dlseder}, {Roques}, {Schanne}, {Shrader},
  {Skinner}, {Strong}, {Sturner}, {Teegarden}, {von Kienlin}, \&
  {Weidenspointner}}]{diehl03}
{Diehl}, R., {Baby}, N., {Beckmann}, V., {et~al.} 2003, \aap, 411, L117

\bibitem[{{Done} {et~al.}(2003){Done}, {Madejski}, {{\. Z}ycki}, \&
  {Greenhill}}]{done03}
{Done}, C., {Madejski}, G.~M., {{\. Z}ycki}, P.~T., \& {Greenhill}, L.~J. 2003,
  \apj, 588, 763

\bibitem[{{Done} {et~al.}(1996){Done}, {Madejski}, \& {Smith}}]{Done96}
{Done}, C., {Madejski}, G.~M., \& {Smith}, D.~A. 1996, \apjl, 463, L63

\bibitem[{{Fanaroff} \& {Riley}(1974)}]{fanaroff74}
{Fanaroff}, B.~L. \& {Riley}, J.~M. 1974, \mnras, 167, 31P

\bibitem[{{Fishman} {et~al.}(1992){Fishman}, {Meegan}, {Wilson}, {Paciesas}, \&
  {Pendleton}}]{fishman92}
{Fishman}, G.~J., {Meegan}, C.~A., {Wilson}, R.~B., {Paciesas}, W.~S., \&
  {Pendleton}, G.~N. 1992, in The Compton Observatory Science Workshop, 26--34

\bibitem[{{Frontera} {et~al.}(1997){Frontera}, {Costa}, {dal Fiume}, {Feroci},
  {Nicastro}, {Orlandini}, {Palazzi}, \& {Zavattini}}]{frontera97}
{Frontera}, F., {Costa}, E., {dal Fiume}, D., {et~al.} 1997, \aaps, 122, 357

\bibitem[{{Gilli} {et~al.}(1999){Gilli}, {Risaliti}, \& {Salvati}}]{gilli99}
{Gilli}, R., {Risaliti}, G., \& {Salvati}, M. 1999, \aap, 347, 424

\bibitem[{{Grandi} {et~al.}(2003){Grandi}, {Fiocchi}, {Perola}, {Urry},
  {Maraschi}, {Massaro}, {Matt}, {Preite-Martinez}, {Steinle}, \&
  {Collmar}}]{grandi03}
{Grandi}, P., {Fiocchi}, M., {Perola}, C.~G., {et~al.} 2003, \apj, 593, 160

\bibitem[{{Guainazzi}(2002)}]{guainazzi02}
{Guainazzi}, M. 2002, \mnras, 329, L13

\bibitem[{{Guainazzi} {et~al.}(1999){Guainazzi}, {Matt}, {Antonelli},
  {Bassani}, {Fabian}, {Maiolino}, {Marconi}, {Fiore}, {Iwasawa}, \&
  {Piro}}]{guainazzi99}
{Guainazzi}, M., {Matt}, G., {Antonelli}, L.~A., {et~al.} 1999, \mnras, 310, 10

\bibitem[{{Guainazzi} {et~al.}(2000){Guainazzi}, {Matt}, {Brandt}, {Antonelli},
  {Barr}, \& {Bassani}}]{guainazzi00}
{Guainazzi}, M., {Matt}, G., {Brandt}, W.~N., {et~al.} 2000, \aap, 356, 463

\bibitem[{{Haardt} \& {Maraschi}(1993)}]{haardt93}
{Haardt}, F. \& {Maraschi}, L. 1993, \apj, 413, 507

\bibitem[{{Hammersley} {et~al.}(1992){Hammersley}, {Ponman}, \&
  {Skinner}}]{hammersley92}
{Hammersley}, A.~P., {Ponman}, T.~J., \& {Skinner}, G.~K. 1992, NIMS, 311, 585

\bibitem[{{Johnson} {et~al.}(1993){Johnson}, {Kinzer}, {Kurfess}, {Strickman},
  {Purcell}, {Grabelsky}, {Ulmer}, {Hillis}, {Jung}, \& {Cameron}}]{johnson93}
{Johnson}, W.~N., {Kinzer}, R.~L., {Kurfess}, J.~D., {et~al.} 1993, \apjs, 86,
  693

\bibitem[{{Jourdain} {et~al.}(1993){Jourdain}, {Bassani}, {Roques}, {Mandrou},
  {Ballet}, {Claret}, {Laurent}, {Lebrun}, {Finogenov}, {Churazov}, {Gilfanov},
  {Sunyaev}, {Dyachkov}, {Khavenson}, {Sukhanov}, \& {Kremnev}}]{jourdain93}
{Jourdain}, E., {Bassani}, L., {Roques}, J.~P., {et~al.} 1993, \apj, 412, 586

\bibitem[{{Lebrun} {et~al.}(2003){Lebrun}, {Leray}, {Lavocat}, {Cr{\' e}tolle},
  {Arqu{\` e}s}, {Blondel}, {Bonnin}, {Bou{\` e}re}, {Cara}, {Chaleil}, {Daly},
  {Desages}, {Dzitko}, {Horeau}, {Laurent}, {Limousin}, {Mathy}, {Mauguen},
  {Meignier}, {Molini{\' e}}, {Poindron}, {Rouger}, {Sauvageon}, \&
  {Tourrette}}]{lebrun03}
{Lebrun}, F., {Leray}, J.~P., {Lavocat}, P., {et~al.} 2003, \aap, 411, L141

\bibitem[{{Leighly} {et~al.}(1999){Leighly}, {Halpern}, {Awaki}, {Cappi},
  {Ueno}, \& {Siebert}}]{leighly99}
{Leighly}, K.~M., {Halpern}, J.~P., {Awaki}, H., {et~al.} 1999, \apj, 522, 209

\bibitem[{{Lund} {et~al.}(2003){Lund}, {Budtz-J{\o}rgensen}, {Westergaard},
  {Brandt}, {Rasmussen}, {Hornstrup}, {Oxborrow}, {Chenevez}, {Jensen},
  {Laursen}, {Andersen}, {Mogensen}, {Rasmussen}, {Om{\o}}, {Pedersen},
  {Polny}, {Andersson}, {Andersson}, {K{\" a}m{\" a}r{\" a}inen}, {Vilhu},
  {Huovelin}, {Maisala}, {Morawski}, {Juchnikowski}, {Costa}, {Feroci},
  {Rubini}, {Rapisarda}, {Morelli}, {Carassiti}, {Frontera}, {Pelliciari},
  {Loffredo}, {Mart{\'{\i}}nez N{\' u}{\~ n}ez}, {Reglero}, {Velasco},
  {Larsson}, {Svensson}, {Zdziarski}, {Castro-Tirado}, {Attina}, {Goria},
  {Giulianelli}, {Cordero}, {Rezazad}, {Schmidt}, {Carli}, {Gomez}, {Jensen},
  {Sarri}, {Tiemon}, {Orr}, {Much}, {Kretschmar}, \& {Schnopper}}]{lund03}
{Lund}, N., {Budtz-J{\o}rgensen}, C., {Westergaard}, N.~J., {et~al.} 2003,
  \aap, 411, L231

\bibitem[{{Madejski} {et~al.}(2000){Madejski}, {{\. Z}ycki}, {Done}, {Valinia},
  {Blanco}, {Rothschild}, \& {Turek}}]{madejski00}
{Madejski}, G., {{\. Z}ycki}, P., {Done}, C., {et~al.} 2000, \apjl, 535, L87

\bibitem[{{Malizia} {et~al.}(2003){Malizia}, {Bassani}, {Stephen}, {Di Cocco},
  {Fiore}, \& {Dean}}]{malizia03}
{Malizia}, A., {Bassani}, L., {Stephen}, J.~B., {et~al.} 2003, \apjl, 589, L17

\bibitem[{{Mas-Hesse} {et~al.}(2003){Mas-Hesse}, {Gim{\' e}nez}, {Culhane},
  {Jamar}, {McBreen}, {Torra}, {Hudec}, {Fabregat}, {Meurs}, {Swings},
  {Alcacera}, {Balado}, {Beiztegui}, {Belenguer}, {Bradley}, {Caballero},
  {Cabo}, {Defise}, {D{\'{\i}}az}, {Domingo}, {Figueras}, {Figueroa}, {Hanlon},
  {Hroch}, {Hudcova}, {Garc{\'{\i}}a}, {Jordan}, {Jordi}, {Kretschmar},
  {Laviada}, {March}, {Mart{\'{\i}}n}, {Mazy}, {Men{\' e}ndez}, {Mi}, {de
  Miguel}, {Mu{\~ n}oz}, {Nolan}, {Olmedo}, {Plesseria}, {Polcar}, {Reina},
  {Renotte}, {Rochus}, {S{\' a}nchez}, {San Mart{\'{\i}}n}, {Smith}, {Soldan},
  {Thomas}, {Tim{\' o}n}, \& {Walton}}]{mas-hesse03}
{Mas-Hesse}, J.~M., {Gim{\' e}nez}, A., {Culhane}, J.~L., {et~al.} 2003, \aap,
  411, L261

\bibitem[{{Masetti} {et~al.}(2004){Masetti}, {Palazzi}, {Bassani}, {Malizia},
  \& {Stephen}}]{masetti04}
{Masetti}, N., {Palazzi}, E., {Bassani}, L., {Malizia}, A., \& {Stephen}, J.~B.
  2004, \aap, 426, L41

\bibitem[{{Matsumoto} {et~al.}(2004){Matsumoto}, {Nava}, {Maddox}, {Leighly},
  {Grupe}, {Awaki}, \& {Ueno}}]{matsumoto04}
{Matsumoto}, C., {Nava}, A., {Maddox}, L.~A., {et~al.} 2004, \apj, 617, 930

\bibitem[{{Matt} {et~al.}(1999){Matt}, {Guainazzi}, {Maiolino}, {Molendi},
  {Perola}, {Antonelli}, {Bassani}, {Brandt}, {Fabian}, {Fiore}, {Iwasawa},
  {Malaguti}, {Marconi}, \& {Poutanen}}]{matt99}
{Matt}, G., {Guainazzi}, M., {Maiolino}, R., {et~al.} 1999, \aap, 341, L39

\bibitem[{{Mattson} \& {Weaver}(2004)}]{mattson04}
{Mattson}, B.~J. \& {Weaver}, K.~A. 2004, \apj, 601, 771

\bibitem[{{Morganti} {et~al.}(1992){Morganti}, {Fosbury}, {Hook}, {Robinson},
  \& {Tsvetanov}}]{morganti92}
{Morganti}, R., {Fosbury}, R.~A.~E., {Hook}, R.~N., {Robinson}, A., \&
  {Tsvetanov}, Z. 1992, \mnras, 256, 1P

\bibitem[{{Paul} {et~al.}(1991){Paul}, {Ballet}, {Cantin}, {Cordier},
  {Goldwurm}, {Lambert}, {Mandrou}, {Chabaud}, {Ehanno}, \& {Lande}}]{paul91}
{Paul}, J., {Ballet}, J., {Cantin}, M., {et~al.} 1991, Advances in Space
  Research, 11, 289

\bibitem[{{Pian} {et~al.}(2005){Pian}, {Foschini}, {Beckmann}, {Sillanp{\"
  a}{\" a}}, {Soldi}, {Tagliaferri}, {Takalo}, {Barr}, {Ghisellini},
  {Malaguti}, {Maraschi}, {Palumbo}, {Treves}, {Courvoisier}, {Di Cocco},
  {Gehrels}, {Giommi}, {Hudec}, {Lindfors}, {Marcowith}, {Nilsson}, {Pasanen},
  {Pursimo}, {Raiteri}, {Savolainen}, {Sikora}, {Tornikoski}, {Tosti}, {T{\"
  u}rler}, {Valtaoja}, {Villata}, \& {Walter}}]{pian05}
{Pian}, E., {Foschini}, L., {Beckmann}, V., {et~al.} 2005, \aap, 429, 427

\bibitem[{{Risaliti}(2002)}]{risaliti02}
{Risaliti}, G. 2002, \aap, 386, 379

\bibitem[{{Risaliti} {et~al.}(2002){Risaliti}, {Elvis}, \&
  {Nicastro}}]{risaliti02b}
{Risaliti}, G., {Elvis}, M., \& {Nicastro}, F. 2002, PASA, 19, 155

\bibitem[{{Rothschild} {et~al.}(1999){Rothschild}, {Band}, {Blanco}, {Gruber},
  {Heindl}, {MacDonald}, {Marsden}, {Jahoda}, {Pierce}, {Madejski}, {Elvis},
  {Schwartz}, {Remillard}, {Zdziarski}, {Done}, \& {Svensson}}]{rothschild99}
{Rothschild}, R.~E., {Band}, D.~L., {Blanco}, P.~R., {et~al.} 1999, \apj, 510,
  651

\bibitem[{{Rothschild} {et~al.}(1998){Rothschild}, {Blanco}, {Gruber},
  {Heindl}, {MacDonald}, {Marsden}, {Pelling}, {Wayne}, \&
  {Hink}}]{rothschild98}
{Rothschild}, R.~E., {Blanco}, P.~R., {Gruber}, D.~E., {et~al.} 1998, \apj,
  496, 538

\bibitem[{{Sazonov} {et~al.}(2004){Sazonov}, {Revnivtsev}, {Lutovinov},
  {Sunyaev}, \& {Grebenev}}]{sazonov04}
{Sazonov}, S.~Y., {Revnivtsev}, M.~G., {Lutovinov}, A.~A., {Sunyaev}, R.~A., \&
  {Grebenev}, S.~A. 2004, \aap, 421, L21

\bibitem[{{Skinner} \& {Connell}(2003)}]{skinner03}
{Skinner}, G. \& {Connell}, P. 2003, \aap, 411, L123

\bibitem[{{Smith} \& {Done}(1996)}]{smith96}
{Smith}, D.~A. \& {Done}, C. 1996, \mnras, 280, 355

\bibitem[{{Steinle} {et~al.}(1998){Steinle}, {Bennett}, {Bloemen}, {Collmar},
  {Diehl}, {Hermsen}, {Lichti}, {Morris}, {Schonfelder}, {Strong}, \&
  {Williams}}]{steinle98}
{Steinle}, H., {Bennett}, K., {Bloemen}, H., {et~al.} 1998, \aap, 330, 97

\bibitem[{{Terrier} {et~al.}(2003){Terrier}, {Lebrun}, {Bazzano}, {Belanger},
  {Bird}, {Blondel}, {David}, {Goldoni}, {Goldwurm}, {Gros}, {Laurent},
  {Malaguti}, {Sauvageon}, {Segreto}, \& {Ubertini}}]{terrier03}
{Terrier}, R., {Lebrun}, F., {Bazzano}, A., {et~al.} 2003, \aap, 411, L167

\bibitem[{{Tran}(1995)}]{tran95}
{Tran}, H.~D. 1995, \apj, 440, 565

\bibitem[{{Treister} \& {Urry}(2005)}]{treister05}
{Treister}, E. \& {Urry}, C.~M. 2005, accepted by ApJ, [astro-ph/0505300]

\bibitem[{{Turner} {et~al.}(1997){Turner}, {George}, {Mushotzky}, \&
  {Nandra}}]{turner97}
{Turner}, T.~J., {George}, I.~M., {Mushotzky}, R.~F., \& {Nandra}, K. 1997,
  \apj, 475, 118

\bibitem[{{Ubertini} {et~al.}(2003){Ubertini}, {Lebrun}, {Di Cocco}, {Bazzano},
  {Bird}, {Broenstad}, {Goldwurm}, {La Rosa}, {Labanti}, {Laurent}, {Mirabel},
  {Quadrini}, {Ramsey}, {Reglero}, {Sabau}, {Sacco}, {Staubert}, {Vigroux},
  {Weisskopf}, \& {Zdziarski}}]{ubertini03}
{Ubertini}, P., {Lebrun}, F., {Di Cocco}, G., {et~al.} 2003, \aap, 411, L131

\bibitem[{{Vedrenne} {et~al.}(2003){Vedrenne}, {Roques}, {Sch{\" o}nfelder},
  {Mandrou}, {Lichti}, {von Kienlin}, {Cordier}, {Schanne}, {Kn{\" o}dlseder},
  {Skinner}, {Jean}, {Sanchez}, {Caraveo}, {Teegarden}, {von Ballmoos},
  {Bouchet}, {Paul}, {Matteson}, {Boggs}, {Wunderer}, {Leleux},
  {Weidenspointner}, {Durouchoux}, {Diehl}, {Strong}, {Cass{\' e}}, {Clair}, \&
  {Andr{\' e}}}]{vedrenne03}
{Vedrenne}, G., {Roques}, J.-P., {Sch{\" o}nfelder}, V., {et~al.} 2003, \aap,
  411, L63

\bibitem[{{Wilkes} {et~al.}(2001){Wilkes}, {Mathur}, {Fiore}, {Antonelli}, \&
  {Nicastro}}]{wilkes01}
{Wilkes}, B.~J., {Mathur}, S., {Fiore}, F., {Antonelli}, A., \& {Nicastro}, F.
  2001, \apj, 549, 248

\bibitem[{{Winkler} {et~al.}(2003{\natexlab{a}}){Winkler}, {Courvoisier}, {Di
  Cocco}, {Gehrels}, {Gim{\' e}nez}, {Grebenev}, {Hermsen}, {Mas-Hesse},
  {Lebrun}, {Lund}, {Palumbo}, {Paul}, {Roques}, {Schnopper}, {Sch{\"
  o}nfelder}, {Sunyaev}, {Teegarden}, {Ubertini}, {Vedrenne}, \&
  {Dean}}]{winkler03a}
{Winkler}, C., {Courvoisier}, T.~J.-L., {Di Cocco}, G., {et~al.}
  2003{\natexlab{a}}, \aap, 411, L1

\bibitem[{{Winkler} {et~al.}(2003{\natexlab{b}}){Winkler}, {Gehrels}, {Sch{\"
  o}nfelder}, {Roques}, {Strong}, {Wunderer}, {Ubertini}, {Lebrun}, {Bazzano},
  {Del Santo}, {Lund}, {Westergaard}, {Beckmann}, {Kretschmar}, \&
  {Mereghetti}}]{winkler03b}
{Winkler}, C., {Gehrels}, N., {Sch{\" o}nfelder}, V., {et~al.}
  2003{\natexlab{b}}, \aap, 411, L349

\bibitem[{{Zdziarski} {et~al.}(1995){Zdziarski}, {Johnson}, {Done}, {Smith}, \&
  {McNaron-Brown}}]{zdziarski95}
{Zdziarski}, A.~A., {Johnson}, W.~N., {Done}, C., {Smith}, D., \&
  {McNaron-Brown}, K. 1995, \apjl, 438, L63

\bibitem[{{Zdziarski} {et~al.}(2000){Zdziarski}, {Poutanen}, \&
  {Johnson}}]{zdziarski00}
{Zdziarski}, A.~A., {Poutanen}, J., \& {Johnson}, W.~N. 2000, \apj, 542, 703

\end{thebibliography}

\end{document}